\documentclass[final,5p,twocolumn,times,sort&compress]{elsarticle}

\usepackage{graphicx}
\usepackage{graphicx}
\usepackage{amsmath,bm}
\usepackage{flushend}
\usepackage{color}
\usepackage{cuted}
\usepackage{widetext}

\def\lsim{\mathrel{\rlap{\lower4pt\hbox{\hskip1pt$\sim$}}
		\raise1pt\hbox{$<$}}}
\def\gsim{\mathrel{\rlap{\lower4pt\hbox{\hskip1pt$\sim$}}
		\raise1pt\hbox{$>$}}}
\def\sqr#1#2{{\vcenter{\vbox{\hrule height.#2pt
				\hbox{\vrule width.#2pt height#1pt \kern#1pt
					\vrule width.#2pt}
				\hrule height.#2pt}}}}
\newcommand{\beq}{\begin{equation}}
	\newcommand{\eeq}{\end{equation}}
\newcommand{\bea}{\begin{eqnarray}}
	\newcommand{\eea}{\end{eqnarray}}
\newcommand{\rf}[1]{(\ref{#1})}

\def\etal{{\it et al.}}

\begin{document}
	
	\begin{frontmatter}
		
		\title{Beta-Decay Spectrum and Lorentz Violation}

		\author{Ralf Lehnert}

		\address{Indiana University Center for Spacetime Symmetries, Bloomington, IN 47405, USA}

\begin{abstract}
	Relativity theory and its underlying Lorentz and CPT invariance represent key principles of physics 
	and therefore require continued experimental scrutiny 
	across the broadest possible range of energy scales and physical systems. 
	Possibilities for tests of these symmetries in precision $\beta$-decay experiments 
	with focus on KATRIN are investigated.
	It is found that countershaded departures from relativity, 
	which represent a particular challenge to detection, 
	are accessible in such experiments. 
	In this context, it is argued that 
	KATRIN would be in an excellent position 
	to conduct the first-ever experimental search for countershaded Lorentz and CPT breaking 
	in neutrinos.
\end{abstract}

\end{frontmatter}

\section{Introduction}

Our current understanding of nature at the fundamental level 
rests on a few cornerstones of physics
that include Lorentz and CPT invariance.
Assessing the integrity of these cornerstones 
is paramount for a robust advancement of physics,
and comprehensive experimental investigations of these symmetries 
therefore represent an indispensable tool in this endeavor.
At the same time,
theoretical approaches seeking to address unanswered questions 
beyond established physics
often accommodate departures from Lorentz and CPT invariance~\cite{strings,ng,foam,varying,other},
providing further impetus 
for probing these foundational spacetime symmetries.

To support such efforts,
a theoretical framework 
for the identification, interpretation, and analysis 
of Lorentz and CPT tests,
the Standard-Model Extension (SME), 
has been developed~\cite{SME}. 
Based on effective field theory,
it incorporates the usual Standard Model 
and General Relativity, 
and allows for perturbative 
but otherwise general departures 
from Lorentz and CPT symmetry 
that are compatible with mild physical requirements. 
For the last two decades,
the SME has served as the basis 
for numerous experimental~\cite{DataTables} and theoretical~\cite{modes,quantum,tools,ChPT,grav,finsler} investigations 
of Lorentz and CPT symmetry 
across a wide range of physical systems, 
and the framework has also found applications 
in adjacent research fields~\cite{otherSME}.

The Lorentz- and CPT-violating contributions
to the SME Lagrangian 
typically predict corrections $E\to E+\delta E_{\rm SME}$ 
to the conventional energy $E$ of the system in question. 
These corrections lead to modifications in dispersion relations, 
bound-state spectra, etc.,
and the resulting effects 
form the basis
for numerous Lorentz and CPT tests. 
For example, 
kinematical studies, 
interferometry, 
spectroscopy, 
polarimetry, 
and matter--antimatter comparisons 
search for such energy shifts.
There are, 
however, 
a class of SME effects 
that can be masked in such studies 
for various reasons. 
Consider,
for instance, 
a nongravitational system,
so that only energy differences $\Delta E = E_{\rm f}-E_{\rm i}$ are measurable.
When $\delta E_{\rm SME}$ gives identical contributions 
to both $E_{\rm f}$ and $E_{\rm i}$ for that system,
$\Delta E$ remains unaffected.
The detection of such SME coefficients 
represents a particular challenge
and may require 
enlarging the system 
together with the identification of more suitable observables.
Such effects have been coined `countershaded'~\cite{Kostelecky:2008in,Diaz:2013saa,Diaz:2013ywa}, 
and they may in principle be harbingers
of thus far unseen, comparatively large relativity violations. 

One of these countershaded effects 
is associated with the SME's $a^{\mu}$-type coefficients.
They are assigned to fermions 
with independent values for each species. 
In flat-spacetime,
these coefficients amount to constant shifts by $a^{\mu}$ 
in the fermion's momentum 
and are known to be undetectable 
in the single-fermion limit. 
Past experimental searches 
have therefore involved multiple species of fermions~\cite{Kostelecky:1999bm,KTeV,FOCUS,BaBar,KLOE,D0,LHCb}
or gravitational backgrounds~\cite{Kostelecky:2008in,Hohensee:2011wt}, 
and they have placed constraints on $a^{\mu}$ coefficients 
for various quark flavors as well as the proton and the electron.
The $a^{\mu}$ coefficients are also known
to affect the phase space in particle decays and collisions.
In this respect, 
the $\beta$ spectrum in particular has been the subject 
of previous theoretical studies~\cite{Diaz:2013saa,Diaz:2013ywa}, 
as its endpoint has been the target 
of various high-precision experimental investigations, 
and continuing efforts along these lines 
are poised for further gains in sensitivity~\cite{Lobashev:1999tp,Troitsk:2011cvm,Aseev:2012zz,Kraus:2004zw,KATRIN:2005fny,KATRIN:2019yun,Aker:2021gma,Formaggio:2011ba}.

The early analyses contained in Refs.~\cite{Diaz:2013saa,Diaz:2013ywa} 
were performed before the construction of the KATRIN experiment was completed. 
The present work revisits
the utility of the $\beta$-decay spectrum for Lorentz and CPT tests
with focus on that particular experiment.
Our analysis extends previous studies valid at the endpoint only 
to the entire spectrum, 
and we update the Lorentz- and CPT-violating phenomenology 
by adequately including KATRIN's $\beta$-electron collection method.
The results presented in this work
are intended to provide the basis 
for the first relativity test 
involving the electron neutrino's $a^{\mu}$ countershaded SME coefficient. 

This paper is structured as follows.
Section~\ref{basics} reviews the model basics and some special features 
that will be exploited in the subsequent analysis.
The general calculation of the $\beta$ spectrum 
including the SME's $a^{\mu}$ coefficients
is presented in Sec.~\ref{Rate}. 
Section~\ref{lo_effects} specializes these results to a KATRIN-like scenario. 
Various explicit expressions and useful results are collected in an appendix. 
Throughout,
we work in flat spacetime with metric signature $(+,-,-,-)$, 
and we adopt natural units $c=\hbar=k_B=\epsilon_0=1$.


\section{Model Basics}
\label{basics}

The SME's countershaded $a^{\mu}$ contributions 
generically couples to spin-$\frac{1}{2}$ fermions $\psi$
via the interaction $\bar{\psi}\,a^{\mu}\gamma_{\mu}\psi$~\cite{SME}.
In the present work,
we focus on the process of tritium $\beta$ decay
\beq
\label{process}
{}^3_1\textrm{T}\to {}^3_2\textrm{He}^{+}+e^{-}+\bar{\nu}_e\,.
\eeq
As this represents a low-energy process
with dynamics primarily governed by the involved two nuclei,
the electron, and the neutrino, 
we may treat all four external legs
effectively as spin-$\frac{1}{2}$ fermions,
as usual~\cite{Simkovic:2007yi}. 
Lorentz and CPT violation of type $a^{\mu}$ in this system 
is then governed by Lagrangian contributions for each of these fermions 
\begin{equation}
	\label{a_type_LV}
	\delta{\cal L}^{a}_{\rm SME}=
	-\bar{\psi}_w\,a_w^{\mu}\gamma_{\mu}\psi_w\,,
\end{equation}
where the species subscript $w\in\{T,H,e,n\}$
labels the tritium, the helium, the electron, and the neutrino,
respectively. 
The SME also contains a plethora of additional types of coefficients 
that may in principle contribute to the process~\rf{process},
but they have often been constrained already in other physical systems
at levels that render them effectively zero 
for our present purposes.

Denoting the physical four-momenta of the free tritium, helium, electron, and antineutrino by, 
respectively,
$k'^{\mu}=(E'_T,\vec{k}\,{}')$, 
$l'^{\mu}=(E'_H,\vec{l}\,{}')$, 
$p'^{\mu}=(E'_e,\vec{p}\,{}')$, 
and $q'^{\mu}=(E'_n,\vec{q}\,{}')$,
the corresponding free-particle dispersion relations are
\begin{align}
\label{physDR}
(k\;\!{}'-a_T)^2 & =M_T^2\,,\hspace{-10mm} & (l\;\!{}'-a_H)^2 & = M_H^2\,,\nonumber\\
(p\;\!{}'-a_e)^2 & =m_e^2\,,\hspace{-10mm} & (q\;\!{}'+a_n)^2 & = m_n^2\,.
\end{align}
Here, 
the mass parameters on the right-hand side of these equations 
have conventional values. 
In these expressions, 
it is understood 
that only positive-energy roots are physical.
Note that the SME coefficient in the neutrino dispersion relation 
enters with a different sign:
as opposed to the other three fermions,
the neutrino participates in the decay~\rf{process}
in its antiparticle~state.

In the absence of interactions,
the four independent spinor redefinitions $\chi_w = \exp(ia_w\!\cdot\! x)\,\psi_w$ 
remove all $a^{\mu}$-type coefficients 
from the Lagrangian, 
so that they would be unobservable~\cite{SME}. 
However,
this reasoning generally fails in the presence of interactions.
Consider,
for example,
a Lagrangian term containing all four fermions, 
as is the case for $\beta$ decay.
Three spinor redefinitions can then be used 
to remove the corresponding $a_w^{\mu}$ from the free fields 
at the cost of generating undesirable spacetime dependences in the interaction. 
These can be neutralized with the fourth spinor redefinition,
which is then no longer independent. 
As a result,
the interaction and three of the fermion species become conventional,
while the remaining one exhibits a net $a^{\mu}$ 
given by a linear combination of the original four $a_w^{\mu}$.
In general,
given interactions 
allow only specific combinations of $a_w^{\mu}$ to become observable. 
This fact is borne out in the analysis below, 
and it represents one of the reasons 
why this type of SME effect is countershaded.

The $\beta$-decay rate in the SME is given by~\cite{Colladay:2001wk}
\begin{equation}
	\label{general_dGamma}
	d\Gamma= 
	\frac{(2\pi)^4}{2 N_T}
	\frac{d^3\vec{l}\,{}'}{(2\pi)^3 2N_H}\frac{d^3\vec{p}\,{}'}{(2\pi)^3 2N_e}\frac{d^3\vec{q}\,{}'}{(2\pi)^3 2N_n}
	|{\cal M}|^2
	\delta^{(4)}
	\,,
\end{equation}
where $N_w$ are possibly momentum-dependent spinor normalizations for the $\psi_w$,
${\cal M}$ is the transition-matrix element for the considered process calculated in the SME,
and $\delta^{(4)}$ is a four-momentum conserving $\delta$ function:
\begin{equation}
	\label{phys_cons}	
	k'^{\mu}=l'^{\mu}+p'^{\mu}+q'^{\mu}\,.
\end{equation}
We note 
that physical observables remain independent of the choices for $N_w$, 
so that convenient spinor normalization factors may be selected. 
For example, 
the usual textbook expression for the decay rate in the Lorentz-symmetric case 
emerges when the $N_w$ are taken to be the $w$ fermion's conventional energies $E_T$, $E_H$, $E_e$, and $E_n$,
and by transforming to the tritium rest frame $E_T\to M_T$.
In the next section,
we exploit this freedom to streamline the Lorentz and CPT-violating $\beta$-decay calculation.


\section{Structure of the Decay Rate}
\label{Rate}

The key ingredient in the decay-rate formula~\rf{general_dGamma} 
is the transition-matrix element ${\cal M}$.
It may be calculated using conventional perturbative techniques,  
albeit with a modified set of Feynman rules incorporating Lorentz- and CPT-breaking SME contributions~\cite{Colladay:2001wk}.
In the present context,
we have disregarded any SME corrections to interaction vertices and the $W$-boson propagator,
so that the $a^{\mu}$-type contributions to ${\cal M}$ 
are entirely encoded in the momentum-space external-leg spinors.
More specifically,
we may start with the conventional expression for the matrix element 
and then simply replace the ordinary $u^{(s)}_w$ and $v^{(s)}_w$ spinors for spin state $s$
with their Lorentz- and CPT-violating SME counterparts $u'^{(s)}_w$ and $v'^{(s)}_w$. 

The coupling to fermions 
of our $a^{\mu}$-type corrections 
is identical to that of $i\partial^{\mu}$. 
As a result,
its principal effect is a shift in the fermions' momenta.
This unique feature 
simplifies the determination of $u'^{(s)}_w$ and $v'^{(s)}_w$ in the SME:
they can be shown to coincide,
up to arbitrary normalization, 
with their conventional Lorentz-symmetric versions
evaluated at `effective momenta' given by
\begin{align}
	\label{eff_mom}	
	k^{\mu} & \equiv k'^{\mu}-a_T^{\mu}\,, &
	l^{\mu} & \equiv l'^{\mu}-a_H^{\mu}\,,\nonumber\\
	p^{\mu} & \equiv p'^{\mu}-a_e^{\mu}\,, & 
	q^{\mu} & \equiv q'^{\mu}+a_n^{\mu}\,.
\end{align}
Expressed in these shifted momenta,
the following explicit relations emerge.
Energy--momentum conservation~\rf{phys_cons} takes the form 
\begin{equation}
	\label{eff_cons}	
	k^{\mu}+a^\mu=l^{\mu}+p^{\mu}+q^{\mu}\quad\textrm{with}\quad a^\mu\equiv a_T^\mu-a_H^\mu-a_e^\mu+a_n^\mu\,.
\end{equation}
We remark in passing 
that Eq.~\rf{physDR} establishes 
that the effective momenta satisfy conventional dispersion relations. 
The corresponding explicit expressions for the SME eigenspinors are
\begin{align}
	\label{spinor_solns}
		u'^{(s)}_T(k') & = \sqrt{\tfrac{N_T(k')}{E_T(k)}}u^{(s)}_T(k)\,, &
		u'^{(s)}_H(l') & = \sqrt{\tfrac{N_H(l')}{E_H(l)}}u^{(s)}_H(l)\,, \nonumber\\
		u'^{(s)}_e(p') & = \sqrt{\tfrac{N_e(p')}{E_e(p)}}u^{(s)}_e(p)\,, &
		v'^{(s)}_n(q') & = \sqrt{\tfrac{N_n(q')}{E_n(q)}}v^{(s)}_n(q)\,.
\end{align}
To arrive at this result,
the standard fermion-energy normalization has been used for the Lorentz-symmetric spinors, 
while the normalization for the SME spinors has been kept arbitrary, 
as in Eq.~\rf{general_dGamma}. 

We are now in a position 
to continue with the determination of ${\cal M}$. 
In the conventional case, 
the dominant contribution to this decay 
is well described by the tree-level contact-interaction amplitude ${\cal M}_0$ 
given by~\cite{Simkovic:2007yi}
\begin{equation}
	\label{interaction}
	i{\cal M}_0=i\frac{G_F V_{ud}\,g_V}{\sqrt{2}}
	\big[\bar{u}_e\gamma^\mu(1-\gamma_5)\,v_n\big]
	\big[\bar{u}_H\gamma_\mu(1+\lambda\gamma_5)\,u_T\big]\,,
\end{equation}
where $G_F$ denotes Fermi's constant,
$V_{ud}$ the ordinary CKM-matrix element, 
and $g_V$ and $g_A$ the usual vector and axial couplings with $\lambda\equiv g_A/g_V \simeq -1.2756\pm0.0013$. 
The spinors are understood to depend 
on the corresponding conventional particle momenta. 
On shell,
these momenta satisfy ordinary energy--momentum conservation.
Nuclear effects and form factors, 
the intermediate $W$ propagator,
bound atomic electrons, 
additional neutrino flavors, 
radiative corrections, etc.\ can in principle also be implemented 
but are subleading.

As per the reasoning presented above, 
we may now proceed in two steps to derive ${\cal M}$ from Eq.~\rf{interaction}. 
First,
we need to replace the ordinary external-leg spinors 
with their SME counterparts.
In a second step, 
we may then employ Eq.~\rf{spinor_solns} 
to express the SME spinors 
by conventional ones with effective momenta.
This yields
\begin{align}
	\label{LVinteraction}
	i{\cal M} = & i\sqrt{\frac{N_T(k')\,N_H(l')\,N_e(p')\,N_n(q')}{E_T(k)\,E_H(l)\,E_e(p)\,E_n(q)}}{\cal M}_0(k,l,p,q)\,,
\end{align}
which in turns leads to
\begin{equation}
	\label{SME_dGamma}
	d\Gamma= 
	\frac{(2\pi)^4}{2 E_T}
	\frac{d^3\vec{l}}{(2\pi)^3 2E_H}\frac{d^3\vec{p}}{(2\pi)^3 2E_e}\frac{d^3\vec{q}}{(2\pi)^3 2E_n}
	|{\cal M}_0|^2
	\delta^{(4)}
	\,,
\end{equation}
where integration variables have been shifted to the effective momenta.
Note that the expression for the physical decay rate~\rf{SME_dGamma} 
is free of unobservable spinor-normalization factors.
We also note 
that the expression for the rate is identical to the conventional textbook result 
with the exception that it contains the effective momenta~\rf{eff_mom}:
although they obey the same dispersion relations as the ordinary momenta,
their conservation equation~\rf{eff_cons} 
enforced by $\delta^{(4)}$ 
involves the Lorentz- and CPT-violating $a^{\mu}$. 
This feature is compatible 
with our analysis of field redefinitions in Sec.~\ref{basics}. 
We finally remark 
that the above reasoning only relies on the properties of the external-leg spinors,
but not on the details of the interaction vertex;
more refined versions for the coupling between the four particles,
such as the inclusion of the Fermi function $F(Z,E_e)$,
can therefore be implemented if desired. 

We continue by specializing 
to the case of unpolarized tritium   
and undetected final spin states, 
as is appropriate for KATRIN.
We may therefore average over the tritium spin 
and sum over the spins of the decay products,
which permits us to replace $|{\cal M}_0|^2$ by
\begin{align}
\label{unpolM}
	\overline{|{\cal M}_0|^2} = 
	&\; (4\,G_F V_{ud}\,g_V)^2\,\Big[(1+\lambda)^2(k\cdot p)(l\cdot q)\nonumber\\
	&\;{}+(1-\lambda)^2(k\cdot q)(l\cdot p)
	-(1-\lambda^2)M_T M_H (p\cdot q)\Big]
\end{align}
in Eq.~\rf{SME_dGamma}. 
This expression is based in the interaction~\rf{interaction} 
prior to implementation of energy--momentum conservation~\rf{eff_cons}.
With this result
and the definition
\begin{align}
	\label{S_def}
	S
	\equiv &\; \big[
	(1+\lambda)^2(k\cdot p)\,\eta_{\mu\nu}+(1-\lambda)^2 k_{\mu}p_{\nu}
	\big] \int\!\frac{d^3q}{E_n(q)}\frac{d^3l}{E_H(l)}\,q^{\mu}l^{\nu}\delta^{(4)}
	\nonumber
	\\
	&{}-(1-\lambda^2)M_T M_H\, p_{\mu} \int\!\frac{d^3q}{E_n(q)}\frac{d^3l}{E_H(l)}\,q^{\mu}\delta^{(4)}\,,
\end{align}
the rate~\rf{SME_dGamma} may be expressed as
\begin{equation}
	\label{spectr1}
	\frac{d\Gamma}{dE_e} =\frac{(G_F V_{ud}\,g_V)^2}{(2\pi)^5 E_T(k)}\,
	|\vec{p}|\int_{\Delta\Omega}d\Omega_e\; S(k,p)\,,
\end{equation}
where $\Delta\Omega$ denotes the solid angle corresponding to the collected electrons, 
and $d\Omega_e$ the solid-angle element associated with $\vec{p}$. 

It is apparent 
that $S$ is a coordinate scalar 
and composed of factors 
transforming covariantly under coordinate changes. 
In part due to our chosen spinor normalization, 
the individual phase-space elements in particular 
are manifestly coordinate invariant. 
The integrals in the first and second line of Eq.~\rf{S_def}
must then take the forms 
$f(w^2)w^{\mu}w^{\nu}+g(w^2)\eta^{\mu\nu}$ and
$h(w^2)w^{\mu}$,
respectively, 
because they depend only on the four-vector 
\beq
\label{r_def}
w^{\mu} \equiv k^{\mu}-p^{\mu}+a^{\mu}\;
\eeq 
contained in $\delta^{(4)}$. 
In this ansatz,
$f$, $g$, and $h$ 
represent coordinate-scalar functions 
to be determined later.
The general structure of $S$ is then
\begin{align}
	\label{Rfgh}
	S(k,p)=&\;
	f(w^2)\big[
	(1+\lambda)^2(k\cdot p)\,w^2+(1-\lambda)^2(k\cdot w)(p\cdot w)
	\big]\nonumber\\
	&\;
	{}+g(w^2)(5+6\lambda+5\lambda^2)(k\cdot p)
	\nonumber\\
	&\;
	{}-h(w^2) M_T M_H(1-\lambda^2)(p\cdot w)\,.
\end{align} 
Since $S(k,p)$ is a coordinate scalar,
it can only depend on $k^{\mu}$ and $p^{\mu}$ 
via $k\cdot p$, $k\cdot a$, and $p\cdot a$. 
In laboratories
in which the tritium decays approximately at rest,
both the tritium and the decay electrons are nonrelativistic.
Then, 
$k^{\mu}$ and $p^{\mu}$ are largely isotropic 
with large zero components 
and small spatial components.
Rotation violation in $S$ 
will therefore generally be suppressed 
relative to isotropic SME effects.

To determine $f$, $g$, and $h$,
we may again exploit coordinate independence 
and boost to a frame in which $w^{\mu}=(w^0,\vec{0})$.
This is always possible
since tritium and the decay electrons 
are both nonrelativistic, 
so that $(k-p)^2 \simeq +|{\cal O}(M^2_T)|$ 
is timelike in the three-momentum range of interest. 
This is then also true for
$(k-p+a)^2 = w^2 \simeq +|{\cal O}(M^2_T)|$ 
because $a^{\mu}$ can only be a small perturbation 
on observational grounds.
In these coordinates,
the integrations simplify. 
We finally obtain
\begin{align}
	\label{fgh}
	f(w^2)
	= &\;\frac{\pi}{3} \frac{1}{w^6}\big[w^4+(M_H^2+m_n^2)w^2-2(M_H^2-m_n^2)^2\big]\times\nonumber\\
	  &\;\sqrt{\big[w^2-(M_H+m_n)^2\big]\big[w^2-(M_H-m_n)^2\big]}\,,\nonumber\\
	g(w^2)
	= &\;\frac{\pi}{6} \frac{1}{w^4}
	\sqrt{\big[w^2-(M_H+m_n)^2\big]\big[w^2-(M_H-m_n)^2\big]}^{\;3}\,,\nonumber\\
	h(w^2)
	= &\;\pi \frac{1}{w^4}\big[w^2-(M_H^2-m_n^2)\big]\times\nonumber\\
	  &\;\sqrt{\big[w^2-(M_H+m_n)^2\big]\big[w^2-(M_H-m_n)^2\big]}\,,
\end{align}
where we have boosted back to the original, general frame. 
Equations~\rf{spectr1}--\rf{fgh} 
provide an explicit characterization 
of the exact tree-level $\beta$-spectrum 
in the presence of $a^{\mu}$-type Lorentz violation. 

The $\beta$ spectrum 
is often presented for tritium decaying at rest, 
a case also relevant for KATRIN.
The analogous situation 
in the present Lorentz-violating context 
is best represented by a configuration 
in which the group velocity 
\begin{equation}
	\label{rest_frame}
	\vec{v}_g 
	=\frac{\partial E'_T}{\partial\; \vec{\!k}{}'}
	=\frac{\vec{\!k}{}'-\vec{a}_T}{\sqrt{M_T^2+(\;\vec{\!k}{}'-\vec{a}_T)^2}}
\end{equation}
of the tritium quantum wave packet is zero.
This requires $\vec{\!k}{}'-\vec{a}_T=\vec{\!k}=\vec{0}$,
i.e., 
the effective (not the physical) three-momentum must vanish.
For such a set-up,
the spectrum~(\ref{spectr1}) simplifies somewhat.


\section{Leading-Order Effects}
\label{lo_effects}

Further insight into $a^{\mu}$-type Lorentz-violation signatures 
in $\beta$ decay 
can be obtained by considering leading-order effects. 
For the special case of tritium decay,
the system is governed by the various mass scales 
listed in Table~\ref{Table1}. 
These may be exploited for the determination of dominant effects.
However,
expansions in $a^{\mu}$, 
in particular close to the endpoint,
require some care. 
Suppose one takes
\begin{equation}
\label{SimpleTaylor}
	\beta(E_e,a^{\mu})\simeq\beta(E_e,0)+\frac{\partial\beta(E_e,0)}{\partial a^{\mu}}a^{\mu}\,.
\end{equation}
While this expansion may be acceptable 
throughout most regions of the spectrum, 
at least two issues arise in the vicinity of the endpoint.
First,
$\beta(E_e,0)$, 
the conventional spectrum,
exhibits a square-root structure 
that yields complex-valued rates 
past the conventional endpoint $E_m$.
If the Lorentz-violating endpoint $E_m^a>E_m$ 
happens to be greater than the conventional one,
expression~(\ref{SimpleTaylor})
becomes invalid past $E_m$ 
and therefore fails to capture the effects of $a^{\mu}$ accurately 
in this region.
Second,
the general square-root dependence of the spectrum close to the endpoint,
which will persist in the SME case (see below),
can lead to a diverging derivative $\frac{\partial\beta(E_e,0)}{\partial a^{\mu}}$ 
as $E_m$ is approached, 
indicating a breakdown of the Taylor expansion~(\ref{SimpleTaylor}).
Note that other pathologies,
such as negative count rates close to the endpoint, 
may also occur.

\begin{table*}
	\begin{center}
		\begin{tabular}{  l  r  c  c }
			\hline\hline\\[-8pt]
			Quantity & Approx.\ Value & Scale (eV) & Rel.\ Suppression \\ [+2pt]
			\hline\\[-8pt]
			$M_T$ & $2.80943\;$GeV & $3\times10^9$ & $\sim 10^{0\phantom{-}}$\\
			$M_H=M$ & $2.80890\;$GeV & $3\times10^9$ & $\sim 10^{0\phantom{-}}$\\
			$\Delta M=M_T-M_H$ & $0.52954\;$MeV & $5\times10^5$ & $\sim 10^{-4}$\\
			$m_e$ & $0.511 00\;$MeV & $5\times10^5$ & $\sim 10^{-4}$\\
			$\Delta E\equiv \Delta M-m_e$ & $18.5378\;$keV & $2\times10^4$ & $\sim 10^{-5}$\\
			$w^2/M\simeq2(E_m-E_e)$ & $0\ldots36\;$keV & $4\times10^4$ & $\sim 10^{-5}$\\
			$|\vec{p}|$ & $0\ldots140\;$keV & $1\times10^5$ & $\sim 10^{-4}$\\
			$m_n$ & $<3\;$eV & $3\times10^0$ & $\sim 10^{-9}$\\
			\hline
		\end{tabular}
		\caption[Table caption text]{Mass scales entering the $\beta$ spectrum of tritium.
		Certain ratios of these may be chosen as expansion parameters.}
		\label{Table1}
		\vskip-12pt
	\end{center}
\vskip-18pt
\end{table*}

To mitigate these issues,
we will set out to cast the leading-order result 
as a correction to $E_m$ 
rather than as a correction to the spectrum~(\ref{SimpleTaylor}).
This should yield a better approximation 
in the vicinity of the endpoint,
where kinematical effects are expected to dominate.
However, 
additional spectral distortions due to $a^{\mu}$ 
may also occur, 
and these effects can be captured separately 
via the usual expansion methods applied to the smoothly varying piece of the spectrum.
For these reasons,
the resulting approximate expression for the spectrum may not be linear in $a^{\mu}$,
but nevertheless only valid at leading order.

As a first step,
a suitable decomposition of $S$ 
into the two relevant pieces 
needs to be determined. 
To this end, 
we write
\beq
\label{S_Decomp}
S=\frac{\pi}{6 w^6}\,R\,P\,,
\eeq
where $R$ contains the square roots in $S$, 
and $P$ is a polynomial in momentum variables.
We take $R$ to be
\begin{align}
	\!\!R(w^2)
	\equiv & \;\sqrt{w^2-(M_H+m_n)^2}\sqrt{w^2-(M_H-m_n)^2}\nonumber\\
	= &\;\sqrt{2M_T(E_m-E_e)+2(k-p)\cdot a+a^2}\;\;\times\nonumber\\
	&\;\sqrt{2M_T(E_m-E_e)+4m_nM_H+2(k-p)\cdot a+a^2}\label{sqrtfactor2}\,,
\end{align}
which is contained in $f$, $g$, and $h$ in Eq.~\rf{fgh},
and hence indeed represents an overall factor in $S$. 
Here,
we have abbreviated the ordinary endpoint energy for stationary tritium as
\begin{equation}
	\label{convEnd}
	E_m\equiv\frac{M_T^2+m_e^2-(M_H+m_n)^2}{2M_T}\,.
\end{equation}
In the usual case,
its first square-root factor
vanishes at $E_e=E_m$ 
determining the endpoint of the beta spectrum.
Its second square-root factor
also exhibits considerable relative variations close to the endpoint,
so that the factor $R$ indeed determines to a large extent 
the shape of the beta spectrum in this region.

In the presence of $a^{\mu}$-type Lorentz violation,
the endpoint of the spectrum is also reached 
when the first square-root term vanishes. 
But the additional contributions to this term 
modify the value of the endpoint $E_m\to E_m^a(\hat{p})$: 
\begin{align}
	\label{newEnd}
	E_m^a(\hat{p})
	\simeq E_m+\left(1-\frac{E_m}{M_T}\right)a^0+\frac{\sqrt{E_m^2-m_e^2}}{M_T}\,\hat{p}\cdot\vec{a}\,,
\end{align}
where $\hat{p}\equiv\vec{p}/|\vec{p}|$, 
and only contributions linear in $a^{\mu}$ have been retained. 
The factor $R$,
and thus the endpoint, 
are therefore indeed anisotropic,
albeit suppressed by $\sqrt{E_m^2-m_e^2}/M_T\sim 10^{-5}$ 
relative to the isotropic shift $a^0$.

The remaining piece $P$ of $S$ given by
\begin{align}
\label{remainR}
	P
	& = 2\,\big[
		w^4+(M_H^2+m_n^2)w^2-2(M_H^2-m_n^2)^2
		\big]\;\times\nonumber\\
	& \phantom{{}={}}
		\big[
		(1+\lambda)^2(k\cdot p)\,w^2+(1-\lambda)^2(k\cdot w)(p\cdot w)
		\big]
		\nonumber\\
	& \phantom{{}={}}
	{}+w^2R^2(5+6\lambda+5\lambda^2)(k\cdot p)\nonumber\\
	& \phantom{{}={}}
	{}-6\,w^2\big[w^2-(M_H^2-m_n^2)\big]M_T M_H(1-\lambda^2)(p\cdot w)
\end{align}
is the desired polynomial factor, 
and thus permits a smooth Taylor expansion.
Using the relation $w^{\mu} = k^{\mu}-p^{\mu}+a^{\mu}$,
a Mandelstam-variable-type reasoning establishes 
that $P$ can be expressed 
as a polynomial in $w^2$, $k\cdot a$, and $p\cdot a$.
At leading order in $a^{\mu}$,
we only need to retain linear powers in $k\cdot a$ and $p\cdot a$.
We also shift the $w^2$ variable to $r^2\equiv w^2-(M_H+m_n)^2$,
which vanishes for $E_e\to E_m$, 
and therefore provides better control of various approximations
in the vicinity of the endpoint. 
This leads to
\begin{equation}
\label{remainR2}
	P=P_r(r^2)+P_k(r^2)\,(k\cdot a)+P_p(r^2)\,(p\cdot a)\,.
\end{equation}
where $P_r(r^2)$, $P_k(r^2)$, and $P_p(r^2)$ 
abbreviate polynomials in $r^2$ 
that emerge in a straightforward way 
from these manipulations. 
Their explicit expressions are given in the appendix. 
At leading order in $a^{\mu}$,
we may also replace $r \to r_0$ 
in both $P_k(r^2)$ and $P_p(r^2)$,
where $r_0^2 \equiv (k-p)^2-(M_H+m_n)^2$ 
no longer contains $a^{\mu}$. 

With the above analysis,
the differential decay rate~\rf{spectr1} 
takes the form
\begin{align}
	\label{spectr2}
	\frac{d\Gamma}{dE_e} & =
	\frac{(2\, G_F V_{ud}\,g_V)^2}{3\,(4\pi)^4 E_T(k)}\,|\vec{p}|\;\times\nonumber\\
	& \phantom{{}={}}
	\int\limits_{\Delta\Omega}\!d\Omega_e\,
	R(r^2)\,\frac{P_r(r^2)+P_k(r_0^2)\,(k\cdot a)+P_p(r_0^2)\,(p\cdot a)}{\big[r^2+(M_H+m_n)^2\big]^3}\,.
\end{align}
The Lorentz-symmetric spectrum emerges for $a^{\mu} \to 0$
and thus $r \to r_0$.
An explicit expression still requires integration 
over the direction of the emitted $\beta$ electrons.
At KATRIN,
these are confined to an acceptance cone,
which we parametrize here by 
the direction $\hat{c}$ of its symmetry axis 
and its opening angle $\kappa$ relative to $\hat{c}$.
As a result of the angular dependence of the endpoint 
generated by $a^{\mu}$, 
situations may arise in which for a given $|\vec{p}|$,
some directions $\hat{p}$ inside the cone 
are part of the spectrum 
while others lie beyond the endpoint 
and are kinematically forbidden. 
The latter momentum directions 
would then have to be excluded from the integration region.
However,
inspection of Eq.~(\ref{newEnd}) reveals 
that this issue is confined to in an  energy band $\pm|\vec{a}|\sqrt{E_m^2-m_e^2}/M_T$
around the median endpoint.
A band width close to the KATRIN data-point spacing of about $1\,$eV
would therefore require an unrealistically large $|\vec{a}|\sim 100\,$keV. 
For this reason,
we may disregard this issue, 
focus on the beta spectrum outside this extremely small band, 
and integrate over the entire acceptance cone.
 
For the actual cone integration,
we opt for an expansion of the integrand into spherical harmonics 
as opposed to a Taylor series.
The expansion coefficients are then determined 
by integrals of the bounded $S(\vec{a}\cdot\vec{p},E_e)$ 
over the unit sphere, 
and they therefore remain finite 
throughout the entire $\beta$ spectrum. 
At electron energies $E_e$ in the vicinity of the endpoint,
both $w^6$ and $P$ in Eq.~\rf{S_Decomp} 
are approximately constant,
and the behavior of the integrand 
is primarily governed by $R$. 
If $E_e$ is several neutrino masses below the endpoint,
$m_n$ is negligible in $R$.
The two square roots in Eq.~\rf{sqrtfactor2} 
become identical
leading to a linear dependence on $\vec{a}\cdot\vec{p}$
so that only the $l=0,1$ terms 
in the spherical expansion survive.
When the endpoint is approached,
a finite $m_n$ generates contributions at higher $l$.
However,
the second square root in Eq.~\rf{sqrtfactor2} 
then approaches $2\sqrt{m_n M_H}$ 
becoming constant at leading order in 
$\vec{a}\cdot\vec{p}/(m_n M_H)$.
Note that this reasoning 
remains valid for $|\vec{a}|$ 
several orders of magnitude larger than $m_n$.
The spherical decomposition of $S(\vec{a}\cdot\vec{p},E_e)$ 
is then governed by the first square root in Eq.~\rf{sqrtfactor2}. 
At the endpoint,
this translates into spherical expansion coefficients 
decreasing as $l^{5/2}$,
which is compatible with uniform convergence. 
In summary,
$S(\vec{a}\cdot\vec{p},E_e)$ is dominated by $l=0,1$ angular behavior 
several multiples of $m_n$ below the endpoint. 
When the endpoint is approached,
contributions at higher $l$ are generated 
but fall off sufficiently rapidly with $l$.

With the expansion of $S(\vec{a}\cdot\vec{p},E_e)$ under control,
the integration over the acceptance cone 
may now be performed:
\begin{equation}
	\label{ang_int}
	\int_{\Delta\Omega}\!\! d\Omega_e\, S(\vec{a}\cdot\vec{p},E_e)
	=\Omega_e\, S(\bar{p}\,\hat{c}\cdot\vec{a},E_e)+{\cal O}(l\ge 2)\,,
\end{equation}
where $\Omega_e\equiv 2 \pi  (1-\cos \kappa)$ 
and $\bar{p}\,\hat{c}\equiv |\vec{p}|\,\hat{c}\cos^2\!\frac{1}{2}\kappa\equiv\vec{\bar{p}}$ may be interpreted  
as an `average' momentum of the electrons inside the acceptance cone. 
In what follows, 
we will drop all higher-order terms ${\cal O}(l\ge 2)$.
It is thus apparent
that the key effect of the integration 
is an overall factor of $\Omega_e$, 
the size of the solid angle subtended by the cone,
as well as a replacement of $\vec{p}$ 
with its average over the cone. 
In particular
\begin{equation}
	\label{w_expression}
	r^2
	\to 2M_T(\bar{E}_m^a-E_e)\equiv \bar{r}^2\,,
\end{equation}
where we have introduced a `mean' endpoint
\beq
\label{mean_endpoint}
\bar{E}_m^a(\hat{c}) \equiv E_m+a^0+M_T^{-1}\sqrt{E_m^2-m_e^2}\,\hat{c}\cdot\vec{a}\,\cos^2\!\tfrac{1}{2}\kappa
\eeq
valid at leading order in $a^{\mu}$,
and where we have linearized in $E_e$ 
by dropping terms of order $E_m(E_m-E_e)/(E_m^2-m_e^2)^{1/2}$.
Physically,
$\bar{E}_m^a(\hat{c})$ represents the endpoint corresponding to the average cone momentum 
and thus depends on the cone direction~$\hat{c}$.

With these considerations,
Eq.~(\ref{spectr2}) becomes in the tritium rest frame:
\begin{align}
	\label{spectr3}
	\frac{d\Gamma}{dE_e} 
	& = \frac{(2\, G_F V_{ud}\,g_V)^2}{3\,(4\pi)^4 M_T}\,
	\frac{|\vec{p}|\,\Omega_e R(\bar{r}^2)}{\big[\bar{r}^2+(M_H+m_n)^2\big]^3}\;\times \nonumber\\
	&\phantom{{}={}} \Big[P_r(\bar{r}^2)+P_k(\bar{r}^2)\,M_T a^0+P_p(\bar{r}^2)\,(E_e a^0-\vec{\bar{p}}\cdot\vec{a})\Big]\,.
\end{align}
Here,
the first term in the numerator's square brackets 
represents the conventional expression,
but with the usual endpoint  $E_m$ 
replaced by the cone-averaged endpoint $\bar{E}^a_m$.
In this term,
$a^{\mu}$ effects leave the shape of the spectrum unchanged,
but shift it horizontally along the $E_e$ axis.
The following two terms 
contain additional corrections 
that also affect the shape of the spectrum:
just like the first term,
they contain the overall square-root factor $R(\bar{r}^2)$
terminating the spectrum at $\bar{E}^a_m$. 
But the $P_k$ and $P_p$ contributions 
also distort the spectrum.

To estimate the relative size of the Lorentz- and CPT-violating 
spectral distortions,
we need to compare $P_r(\bar{r}^2)$, $P_k(\bar{r}^2)$, and $P_p(\bar{r}^2)$ 
across the spectrum. 
Working with the approximations~(\ref{pwappr}), (\ref{pkwappr}), and (\ref{ppwappr})
and taking only terms up to $\bar{r}^2$ into consideration,
we have:
\begin{align}
\label{dist}
	\frac{P_k(\bar{r}^2)M+P_p(\bar{r}^2)E_e}{P_r(\bar{r}^2)}\,a^0
	&\simeq\frac{a^0}{\Delta M}\,,\\
	-\frac{P_p(\bar{r}^2)\,\bar{p}}{P_r(\bar{r}^2)}\,(\hat{c}\cdot \vec{a})
	&\simeq\frac{(1+\lambda)^2}{1+3\lambda^2}\,\frac{\bar{p}}{M\,\Delta M}\,(\hat{c}\cdot \vec{a})\,.
\end{align}
The first equation represents an energy-independent, isotropic scaling of the beta spectrum. 
To arrive at this expression,
we dropped a term of order $E_e/M\simeq{\cal O}(10^{-4})$.
The second line captures the leading anisotropic effects.
Its only energy dependence arises 
via the average cone momentum $\vec{\bar{p}}$
and therefore vanishes at the beginning of the spectrum 
and is maximal at the endpoint; 
its dependence on the cone direction $\hat{c}$ is apparent. 
Note, however,
that the anisotropic effect is suppressed by at least $6\times 10^{-6}$
relative to the isotropic correction.
 
The above spectral corrections 
are not expressed in terms of the physical, conserved $p'^{\mu}$ momentum,
but rather in terms the effective $p^{\mu}=p'^\mu-a_e^\mu$.
To match our predictions to the actual experimental set-up,
we still need to consider the interplay of $p^{\mu}$ 
with the KATRIN spectrometer. 
Once the electron escapes the decay region,
the relevant model describing the electron--spectrometer system 
is single-flavor quantum electrodynamics 
supplemented by a Lorentz-violating correction 
of the form (\ref{a_type_LV}) for the electron.
For this reduced system, 
the field redefinition $\chi_{e} = \exp(ia_e\cdot x)\psi_{e}$ 
removes $a_e^\mu$ from the Lagrangian,
so that the electron--spectrometer system 
remains unaffected by this type of Lorentz violation.
In particular,
the physics of the MAC-E filters 
and the relation between $p^\mu$ 
and the retarding voltage 
are conventional.

The Lorentz- and CPT-violating $a^{\mu}$ represents a 4-vector.
Meaningful measurement results 
and the comparisons between different experiments 
therefore benefit from a standard choice of inertial coordinate system 
in which the cartesian components of $a^{\mu}$ are taken as constant.
The Sun-centered celestial equatorial frame (SCCEF)
has become the canonical coordinate system for this purpose~\cite{Kostelecky:2002hh}.
The motion of a terrestrial laboratory 
relative to the SCCEF 
is dominated by three transformations. 
The leading effect arises from a change in orientation 
due to the earth's rotation about its axis. 
This rotation is characterized 
by the earth's sidereal frequency $\omega_{\oplus}\simeq 2\pi/(23\,\textrm{h}\,56\,\textrm{min})$. 
The motion of the earth around the sun 
is associated with the second transformation, 
a boost governed by the orbital speed $\beta_{\oplus}\simeq 9.9\times10^{-5}$ and frequency $\Omega_{\oplus}\simeq 2\pi\,(\textrm{year})^{-1}$
of the earth. 
The third transformation 
originates from the circular motion of the laboratory 
around the earth's axis. 
It is controlled 
by the tangential speed $\beta_L=\omega_{\oplus} r_{\oplus}\sin\chi$,
where $r_{\oplus}$ is the earth's radius
and $\chi$ the laboratory's colatitude.

To fix the the transformation matrix 
from the SCCEF to the experimental setup at earth,
a laboratory frame needs to be selected:
we choose the $x$ axis to point south,
the $y$ axis east,
and the $z$ axis to the local zenith. 
For this choice, 
the explicit form of the transformation matrix $\Lambda^{\mu}{}_{\nu}$, 
where $\mu$ is a laboratory-frame index and $\nu$ a SCCEF index,
may be inferred from Ref.~\cite{Kostelecky:2018fmc} 
and is given here in the appendix as Eq.~\rf{LT}. 
With respect to these laboratory coordinates,
we parametrize the cone direction 
as
\beq\label{conedirection}
\hat{c}=(\sin\alpha\cos\beta,\sin\alpha\sin\beta,\cos\alpha)\,.
\eeq
Here,
$\alpha$ and $\beta$ denote the usual polar and azimuthal angles 
in spherical coordinates 
based on our choice of laboratory axes. 

The rate formula~\rf{spectr3} 
and the definitions~\rf{sqrtfactor2}, 
\rf{w_expression}, 
\rf{mean_endpoint},
\rf{pr},
\rf{pk},
and \rf{pp}
of its individual pieces 
together with the transformation~\rf{LT} 
and the acceptance-cone parametrization~\rf{conedirection} 
provide a general, explicit form 
of the $\beta$-decay rate of tritium 
in the presence of countershaded $a^{\mu}$-type Lorentz and CPT violation.
For KATRIN,
the cone direction is horizontal $\alpha=90^\circ$, 
and current data sets cover time periods 
for which the orbital motion of the earth may be disregarded $\beta_{\oplus}\to 0$. 
This yields 
for $a^{\mu}$ 
in terms of its SCCEF components $a^T$, $a^X$, $a^Y$, $a^Z$
\beq
a^{\mu}=
\left(
\begin{array}{c}
	a^T+a^X \beta _L \sin \omega_{\oplus} t -a^Y\beta _L \cos \omega_{\oplus} t  \\
	a^X \cos \chi \cos \omega_{\oplus} t +a^Y \cos \chi \sin \omega_{\oplus} t - a^Z \sin \chi  \\
	- a^T \beta _L -a^X \sin \omega_{\oplus} t + a^Y \cos \omega_{\oplus} t  \\
	a^X \cos \omega_{\oplus} t \sin \chi +a^Y \sin \chi \sin \omega_{\oplus} t +a^Z \cos \chi  \\
\end{array}
\right),
\eeq
where $t$ denotes sidereal time. 
A sample prediction is then
that the mean KATRIN endpoint exhibits sidereal variations:
\begin{align}\label{Kend}
	\bar{E}_m^a&=
	E_m+a^T-C (a^T \beta _L \sin \beta + a^Z \cos \beta \sin \chi)\nonumber\\
	& \phantom{{}={}}
	{}+(C a^Y \sin \beta+C a^X \cos \beta \cos \chi -a^Y \beta _L )\cos \omega_{\oplus} t\nonumber\\
	& \phantom{{}={}}
	{}+(C a^Y \cos \beta \cos \chi-C a^X \sin \beta+a^X \beta _L)\sin \omega_{\oplus} t\,,\nonumber\\
	&=
	E_m+\tfrac{1}{\sqrt{4\pi}}(\sqrt{3}\, a_{10}\, C \cos \beta \sin \chi-a_{00}\, C\, \beta _L \sin \beta +a_{00})\nonumber\\
	&\phantom{{}={}}
	+\sqrt{\tfrac{3}{8 \pi }}  \big[C \cos \beta \cos \chi-i \left(\beta _L-C \sin \beta\right)\big]\,a_{11} e^{i \omega_{\oplus} t}\nonumber\\ 
	&\phantom{{}={}}
	+\sqrt{\tfrac{3}{8 \pi }}  \big[C \cos \beta \cos \chi+i \left(\beta _L-C \sin \beta\right)\big]\,a_{11}^* e^{-i \omega_{\oplus} t}\,.
\end{align}
Here, the second equality 
presents the result with $a^{\mu}$ decomposed into spherical components~\cite{SME},
and where we have set
\beq\label{Cdef}
C=M_T^{-1}\sqrt{E_m^2-m_e^2}\,\cos^2\!\tfrac{1}{2}\kappa\,.
\eeq
Additional predictions,
such as both constant and sidereal-time dependent spectrum distortions, 
may be determined analogously.

To obtain a rough estimate of KATRIN's reach for $a^{\mu}$ measurements,
we take 
$\beta=164^\circ$,
$\kappa=50^\circ$,
and $\chi=41^\circ$. 
Since the experiment is designed to detect sub-eV neutrino masses, 
a $1\,$eV sensitivity to sidereal endpoint variations 
seems plausible.
These numbers then translate into measurements 
of $a^X$ and $a^Y$ at the order of $10^{-5}\,$GeV. 
The difference between the measured and the theoretical value of the tritium endpoint 
is also accessible at the $1\,$eV level~\cite{KATRIN:2019yun}. 
This should permit a measurement of the coefficient combination $a^T+2.6\times10^{-5}a^Z$ 
at the order of $10^{-9}\,$GeV.
Such analyses would represent the world's first experimental results 
for the $a^{\mu}$ Lorentz- and CPT-violating SME coefficient 
defined in Eq.~\rf{eff_cons}.

Further opportunities for the exploration of $a^{\mu}$ parameter space with KATRIN 
are possible
and could yield both improved sensitivities 
and access to additional observables. 
For example,
the search for 
sidereal variations in the tritium $\beta$ spectrum 
further below the endpoint,
where higher count rates are expected to yield substantially enhanced statistics, 
may lead to significantly improved measurements of $a^X$ and $a^Y$.
Likewise,
data campaigns covering time spans of several months 
would open an avenue 
to investigate effects arising from the boost transformation 
due to the orbital motion of the earth. 
Count rates would then be predicted to vary with frequencies 
composed of both $\omega_{\oplus}$ and $\Omega_{\oplus}$ 
allowing for the measurement of additional amplitudes, 
and thus further $a^{\mu}$-component combinations.


\section*{Acknowledgments}

The author is grateful for discussions with S.~Mertens 
and J.~Wickles. 
This work was supported in part
by the Indiana University Center for Spacetime Symmetries.


\appendix
\section{Explicit Expressions for Key Quantities}
\label{app}

This appendix collects various explicit expressions 
for quantities appearing in the main text.
They are indispensable for a complete
characterization of tritium $\beta$ decay 
in the presence of $a^{\mu}$,
but have been relegated to this appendix 
due to their complexity.
One set of these are $P_r$, $P_k$, and $P_p$:\hfill
\vspace*{-12pt}
\begin{widetext}
	\begin{align}
		P_r(r^2)
		& =\tfrac{1}{2}
		\Big\{
		6 M_H m_n\big[M_H+m_n\big]^2
		+\big[3M_H^2+4M_H m_n+3m_n^2\big]r^2
		+r^4
		\Big\}
		\Big\{
		\big[M_T^2-m_e^2\big]^2
		\big[1-\lambda\big]^2
		\nonumber\\
		&\phantom{{}={}}{}-\big[(M_H+m_n)^2+r^2\big]^2
		\big[1-\lambda\big]^2
		+2\big[
		M_T^2+m_e^2-(M_H+m_n)^2-r^2
		\big]
		\big[
		(M_H+m_n)^2+r^2
		\big]
		\big[1+\lambda\big]^2
		\Big\}\nonumber\\
		&\phantom{{}={}}{}+\tfrac{1}{2}r^2
		\big[
		M_T^2+m_e^2-(M_H+m_n)^2-r^2
		\big]
		\big[
		(M_H+m_n)^2+r^2
		\big]
		\big[
		4M_H m_n+r^2
		\big]
		\big[
		5+ 6 \lambda + 5 \lambda ^2
		\big]
		\nonumber\\
		&\phantom{{}={}}{}-3M_T M_H
		\big[
		M_T^2-m_e^2-(M_H+m_n)^2-r^2
		\big]
		\big[
		(M_H+m_n)^2+r^2
		\big]
		\big[
		2m_n(M_H+m_n)+r^2
		\big]
		\big[
		1- \lambda ^2
		\big]\,,\label{pr}\\[6pt] 
		&\simeq
		6(1+3\lambda^2)M^5\Delta M(2m_n M+r^2)
		-3(1+3\lambda^2)M^4r^4
		\,,\label{pwappr}\\[6pt]
		P_k(r^2)
		& = \Big\{
		(M_H+m_n)^2+r^2
		\Big\}\Big\{r^2
		\big[
		4M_H m_n+r^2
		\big]
		\big[
		5+ 6 \lambda + 5 \lambda ^2
		\big]
		-6M_T M_H
		\big[
		2m_n(M_H+m_n)+r^2
		\big]
		\big[
		1- \lambda ^2
		\big]\Big\}\nonumber\\
		& \phantom{{}={}}{}
		+\Big\{
		6 M_H m_n\big[M_H+m_n\big]^2
		+\big[3M_H^2+4M_H m_n+3m_n^2\big]r^2
		+r^4
		\Big\}
		\nonumber\\
		&\phantom{{}={}}{}
		\times\Big\{
		\big[
		M_T^2-m_e^2+(M_H+m_n)^2+r^2
		\big]
		\big[1-\lambda\big]^2	
		+2\big[
		(M_H+m_n)^2+r^2
		\big]
		\big[1+\lambda\big]^2
		\Big\}\,,\label{pk}\\[6pt]
		&\simeq
		6M^4(1+3\lambda^2)(2m_n M+r^2)
		\,,\label{pkwappr}\\[6pt]
		P_p(r^2)
		& = -\big[
		(M_H+m_n)^2+r^2
		\big]r^2
		\big[
		4M_H m_n+r^2
		\big]
		\big[
		5+ 6 \lambda + 5 \lambda ^2
		\big]\nonumber\\
		& \phantom{{}={}}{}
		+\Big\{
		6 M_H m_n\big[M_H+m_n\big]^2
		+\big[3M_H^2+4M_H m_n+3m_n^2\big]r^2
		+r^4
		\Big\}
		\nonumber\\
		&\phantom{{}={}}{}
		\times\Big\{
		\big[
		M_T^2-m_e^2-(M_H+m_n)^2-r^2
		\big]
		\big[1-\lambda\big]^2	
		-2\big[
		(M_H+m_n)^2+r^2
		\big]
		\big[1+\lambda\big]^2
		\Big\}\,,\label{pp}\\[6pt]
		&\simeq
		6M^3(2m_n M+r^2)\big[(1-\lambda)^2\Delta M-(1+\lambda)^2M\big]
		-16(1+\lambda+\lambda^2)M^2 r^4\,.\label{ppwappr}\hfill
	\end{align}
\end{widetext} \vspace*{-12pt}\\
\noindent 
We have also listed approximations for $P_r$, $P_k$, and $P_p$ 
with sub-percent level precision.
The approximations have been expressed in terms of
$M\equiv M_H$ and $\Delta M\equiv M_T-M_H$.
Table~\ref{Table1} gives approximate values 
of some physical parameters 
entering the $\beta$-spectrum expression; 
the leading-order results Eqs.~(\ref{pwappr}), (\ref{pkwappr}), and~(\ref{ppwappr})
are  based on the suppression factors 
shown in its last column.

The other relation \hfill
\vspace*{-12pt}
\begin{widetext}
\begin{equation}\label{LT}
	\Lambda^{\mu}{}_{\nu}=
	\left(
	\begin{array}{cccc}
		1 & \beta_L s_{\omega }-\beta_{\oplus} s_{\Omega } & \beta_{\oplus} c_{\eta } c_{\Omega }-\beta_L c_{\omega } & \beta_{\oplus} c_{\Omega } s_{\eta } \\
		\beta_{\oplus} (c_{\Omega } c_{\eta } c_{\chi } s_{\omega }-c_{\Omega } s_{\eta } s_{\chi }-c_{\chi } c_{\omega } s_{\Omega }) & c_{\chi } c_{\omega } & c_{\chi }
		s_{\omega } & -s_{\chi } \\
		\beta_{\oplus} c_{\eta } c_{\omega } c_{\Omega }+\beta_{\oplus} s_{\omega } s_{\Omega }-\beta_L & -s_{\omega } & c_{\omega } & 0 \\
		\beta_{\oplus} (c_{\chi } c_{\Omega } s_{\eta }+s_{\chi } c_{\eta } c_{\Omega } s_{\omega }-s_{\chi } c_{\omega } s_{\Omega }) & c_{\omega } s_{\chi } & s_{\chi
		} s_{\omega } & c_{\chi } \\
	\end{array}
	\right)
\end{equation}
\end{widetext}\vspace*{-12pt}\\
pertains to the change of frames 
from SCCEF 
to laboratory coordinates. 
Here,
we have abbreviated the sine and cosine functions with $s$ and $c$, 
respectively.
The subscripts on these abbreviations designate the function's argument. 
In particular,
$\omega$ denotes the phase $\omega_{\oplus}t$ 
and $\Omega$ the phase $\Omega_{\oplus}t$,
where $t$ represents the sidereal time.

\end{document}